\documentstyle [12pt] {article}
\begin{document}
\setcounter{page}{1}
\def\theequation{\arabic{section}.\arabic{equation}}
\def\theequation{\thesection.\arabic{equation}}
\setcounter{section}{0}

\title{On the $\Lambda^+_c \to p + K^- + \pi^+$ decay}

\author{Ya. A. Berdnikov~\thanks{E--mail: berdnikov@twonet.stu.neva.ru,
State Technical University, Department of Nuclear
Physics, 195251 St. Petersburg, Russian Federation} ,  A. N.
Ivanov~\thanks{E--mail: ivanov@kph.tuwien.ac.at, Tel.: +43--1--58801--5598,
Fax: +43--1--5864203}~{\small$^{\S}$} ,
V. F. Kosmach~\thanks{State Technical University, Head of the Department of
Nuclear
Physics, 195251 St. Petersburg, Russian Federation} ,\\
N. I. Troitskaya~\thanks{Permanent Address:
State Technical University, Department of Nuclear
Physics, 195251 St. Petersburg, Russian Federation}}

\date{}

\maketitle

\begin{center}
{\it Institut f\"ur Kernphysik, Technische Universit\"at Wien, \\
Wiedner Hauptstr. 8-10, A-1040 Vienna, Austria}
\end{center}

\vskip1.0truecm
\begin{center}
\begin{abstract}
The proton energy  spectrum and the angular distribution of the probability
of the $\Lambda^+_c \to p \,+ \,K^- + \,\pi^+$ decay for the polarized
$\Lambda^+_c$ and the unpolarized proton are calculated in the effective
quark model with chiral $U(3)\times U(3)$ symmetry incorporating Heavy
Quark Effective theory (HQET) and Chiral perturbation theory at the quark
level (CHPT)$_q$. The application of the obtained result to the analysis of
the polarization of the $\Lambda^+_c$ produced in the processes of photo
and hadroproduction is discussed. We draw the similarity between the
measurements of the polarization of the $\Lambda^+_c$ in the $\Lambda^+_c
\to p \,+ \,K^- + \,\pi^+$ decay and the $\mu^-$--meson in the $\mu^- \to
{\rm e}^- + \bar{\nu}_{\rm e} + \nu_{\mu}$ decay.
\end{abstract}
\end{center}

\newpage

\section{Introduction}
\setcounter{equation}{0}

It is rather likely that in the reactions of photo and hadroproduction the
charmed baryon $\Lambda^+_c$ is produced polarized [1]. The restoration of
the $\Lambda^+_c$ polarization by means of the investigation of the decay
products  should clarify the mechanism of the charmed baryon production at
high energies.

The most favourable mode of the $\Lambda^+_c$ decays to be detected
experimentally is $\Lambda^+_c \to p + K^- + \pi^+$. The experimental
probability of this mode equals to $B(\Lambda^+_c \to p  K^-  \pi^+)_{\exp}
= (5.0\pm 1.3)\,\%$ [2]. However, from the theoretical point of view this
mode is the most difficult case due to the impossibility to factorize the
baryonic and mesonic degrees of freedom for the computation of the matrix
element of the transition [1].

The theoretical investigation of non--leptonic decays  of charmed baryons
without factorization of baryonic and mesonic degrees of freedom can be
carried out in the effective quark model with chiral $U(3)\times U(3)$
symmetry incorporating Heavy Quark Effective Theory (HQET) [3,4] and Chiral
perturbation theory at the quark level (CHPT)$_q$ [5]. The effective quark
model with chiral $U(3)\times U(3)$ symmetry is motivated by the
low--energy effective QCD with a linearly rising interquark potential
responsible for a quark confinement [6]. The application of this model to
the description of the low--energy properties of charmed mesons: mass
spectra [7], coupling constants [7--9], the form factors of the
semileptonic decays [10] and the probabilities of the decays [7,11] gave
the results agreeing good with experimental data.

Recently [12] the effective quark model with chiral $U(3)\times U(3)$
symmetry has been extended by the inclusion of the low--lying baryon octet
and decuplet coupled with the three--quark currents. Due to the  dynamics
of strong low--energy interactions caused by a linearly rising interquark
potential there has been shown [6] that: (i) baryons are the three--quark
states [13] and do not contain any bound diquark states, then (ii) the
spinorial structure of the three--quark currents is defined as the products
of the axial--vector diquark densities $\bar{q^c}_i(x)\gamma^{\mu}q_j(x)$
and a quark field $q_k(x)$ transforming under $SU(3)_f\times SU(3)_c$ group
like $(\underline{6}_f,\tilde{\underline{3}}_c)$ and
$(\underline{3}_f,\underline{3}_c)$ multiplets, respectively, where $i,j$
and $k$ are the colour indices running through $i=1,2,3$ and $q = u,d$ or
$s$ quark field. This agrees with the structure of the three--quark
currents used for the investigation of the properties of baryons within QCD
sum rules approach [14]. The fixed structure of the three--quark currents
allows to describe all variety of low--energy interactions of baryon octet
and decuplet in terms of the phenomenological coupling constant $g_{\rm B}$
describing the coupling of the baryon octet and decuplet with the
three--quark currents [12]:
\begin{eqnarray}\label{label1.1}
{\cal L}_{\rm int}(x) = \frac{1}{\sqrt{2}}g_{\rm
B}\,\bar{B}_{\underline{8}}(x)\,\eta_{\underline{8}}(x) + g_{\rm
B}\,\bar{B}_{\underline{10}}(x)\,\eta_{ \underline{10}}(x) + {\rm h.c.},
\end{eqnarray}
where $B_{\underline{8}}(x) = (\psi_{\rm p}(x),\ldots)$ and
$B_{\underline{10}}(x) = (\psi_{\Delta^{++}},\ldots)$ are the fields of the
baryon octet and decuplet, respectively, and $\eta_{\underline{8}}(x) = ( -
\,\varepsilon^{ijk}\,[\bar{u^c}_i(x) \gamma^{\mu}
u_j(x)]\gamma_{\mu}\gamma^5 d_k(x), \ldots)$ and $\eta_{\underline{10}}(x)
= (\varepsilon^{ijk}\,[\bar{u^c}_i(x) \gamma^{\mu} u_j(x)] u_k(x), \ldots)$
are the three--quark currents. The numerical value of $g_{\rm B}$,
calculated in terms of the coupling constant $g_{\rm \pi NN} = 13.4$ of the
${\rm \pi NN}$ interaction,  has been found equal to $g_{\rm B} =1.34\times
10^{-4}\,{\rm MeV}$ [12]. The coupling constants $g_{\rm \pi N \Delta}$
and $g_{\rm \gamma N \Delta}$ of the ${\rm \pi N \Delta}$ and ${\rm \gamma
N \Delta}$ interactions relative to the coupling constant $g_{\rm \pi NN}$
and the $\sigma_{\rm \pi N}$--term of the low--energy ${\rm \pi
N}$--scattering have been calculated in good agreement with the
experimental data and other phenomenological approaches based on QCD [12].

In this paper we apply the effective quark model with chiral $U(3)\times
U(3)$ symmetry to the calculation of the proton energy spectrum of the
$\Lambda^+_c \to p + K^- + \pi^+$ decay of the polarized $\Lambda^+_c$ and
the analysis of the restoration of the $\Lambda^+_c$ polarization in the
dependence of the energies and momenta of the decay products. For the
analysis of the $\Lambda^+_c \to p + K^- + \pi^+$ decay at the quark level
we assume that the $\Lambda^+_c$ is a three--quark state coupled with the
three--quark current $\eta_{\Lambda^+_c}(x) = - \varepsilon^{ijk}
[\bar{u^c}_i(x) \gamma^{\mu} d_j(x)] \gamma_{\mu} \gamma^5 c_k(x)$ defined
as the product of the axial--vector light--diquark density $\bar{u^c}_i(x)
\gamma^{\mu} d_j(x)$ and the $c$--quark field $ c_k(x)$.

The paper is organized as follows. In Sect.$\,$2 we discuss the effective
low--energy Lagrangian describing non--leptonic decays of charmed hadrons
and reduce the calculation of the amplitude of the $\Lambda^+_c \to p + K^-
+ \pi^+$ decay to the calculation of the matrix element of the low--energy
transition $\Lambda^+_c \to p + K^-$. In Sect.$\,$3 we calculate the matrix
element of the low--energy transition $\Lambda^+_c \to p + K^-$. In
Sect.$\,$4 we calculate the proton energy spectrum of the $\Lambda^+_c \to
p + K^- + \pi^+$ decay for the polarized $\Lambda^+_c$ and the unpolarized
proton. In Sect.$\,$5 we calculate the probability of the $\Lambda^+_c \to
p + \bar{K}^0$ decay relative to the probability of the $\Lambda^+_c \to p
+ K^- + \pi^+$ decay. The theoretical result on the ratio of the
probabilities agrees well with the experimental data. In the Conclusion we
discuss the obtained results. In the Appendix we calculate the momentum
integrals describing the the matrix elements of the low--energy transitions
$\Lambda^+_c \to p + K^-$ and  $\Lambda^+_c \to p + \bar{K}^0$.

\section{Effective Lagrangian for non--leptonic decays of charmed hadrons}
\setcounter{equation}{0}

The effective low--energy Lagrangian responsible for non--leptonic decays
of charmed hadrons reads [15]
\begin{eqnarray}\label{label2.1}
{\cal L}_{\rm eff}(x) &=& -\frac{G_F}{\sqrt{2}}\,V^*_{c s}\,V_{u d}\,
 \Big\{C_1(\mu)\,[\bar{s}(x)\,\gamma^{\mu}(1-\gamma^5)\,c(x)]\,[\bar{u}(x)
 \,\gamma_{\mu} (1-\gamma^5)\,d(x)]\nonumber\\
&& \hspace{0.7in} +
C_2(\mu)\,[\bar{u}(x)\,\gamma^{\mu}(1-\gamma^5)\,c(x)]\,[\bar{s}(x)\,\gamma_{\mu
}(1-\gamma^5)\,d(x)]\Big\},
\end{eqnarray}
where $G_F=1.166\times 10^{-5}\;{\rm GeV}^{-2}$ is the Fermi weak constant,
$V^*_{c s}$ and $V_{u d}$ are the elements of the CKM--mixing matrix,
$C_i(\mu)\,(i=1,2)$  are the Wilson coefficients caused by the strong
quark--gluon interactions at scales $p > \mu$ (short--distance
contributions), where $\mu$ is a normalization scale. In the absence of
quark--gluon interactions the coefficients $C_1(\mu) $ and $C_2(\mu)$ do
not depend on $\mu$ and amount to $C_1 = 1$ and $C_2 = 0$. In (CHPT)$_q$ we
should identify $\mu$ with the scale of spontaneous breaking of chiral
symmetry (SB$\chi$S) $\Lambda_{\chi}= 940\,{\rm MeV}$ [5--11], i.e. $\mu =
\Lambda_{\chi}= 940\,{\rm MeV}$. Therefore, below we would deal with
$C_1(\Lambda_{\chi})$ and $C_2(\Lambda_{\chi})$. The contribution of strong
low--energy interactions at scales $p \le \mu=\Lambda_{\chi}$
(long--distance contributions) is described by (CHPT)$_q$ in terms of
constituent quark loop diagrams, where the  momenta of virtual quarks are
restricted from above by the SB$\chi$S scale $\Lambda_{\chi}$ [5--12].

The structure of the term proportional to $C_2(\Lambda_{\chi})$ can be
reduced to the structure of the first one by means of the Fierz
transformation [15]. The resultant coefficient of the first term would
contain $C_2(\Lambda_{\chi})$ in the form $C_2(\Lambda_{\chi})/N$, where
$N=3$ is the number of quark colours. Thus, the effective low--energy
Lagrangian responsible for the $\Lambda^+_c \to p + K^- + \pi^+$ decay can
be taken in the form
\begin{eqnarray}\label{label2.2}
{\cal L}_{\rm eff}(x) = -\frac{G_F}{\sqrt{2}}\,V^*_{c s}\,V_{u
d}\,\bar{C}_1(\Lambda_{\chi})\,[\bar{s}(x)\,\gamma_{\mu}(1-\gamma^5)\,c(x)]\,[\b
ar{u}(x)\,\gamma^{\mu}(1-\gamma^5)\,d(x)],
\end{eqnarray}
where $\bar{C}_1(\Lambda_{\chi}) = C_1(\Lambda_{\chi}) +
C_2(\Lambda_{\chi})/3$. The amplitude of the $\Lambda^+_c \to p + K^- +
\pi^+$ decay is then defined
\begin{eqnarray}\label{label2.3}
\frac{\displaystyle{\cal M}(\Lambda^+_c(Q) \to p(q)
K^-(q_-)\pi^+(q_+))}{\displaystyle \sqrt{2E_{\Lambda^+_c}V\,2E_p
V\,2E_{K^-} V\,2E_{\pi^+}V}}= <p(q) K^-(q_-)\pi^+(q_+)|{\cal L}_{\rm
eff}(0)|\Lambda^+_c(Q)>,
\end{eqnarray}
where $E_i\,(i=\Lambda^+_c,p,K^-,\pi^+)$ are the energies of the
$\Lambda^+_c$,  the proton and the $K^-$ and $\pi^+$ mesons, respectively,
and $V$ is the normalization volume.

In (CHPT)$_q$ at the tree--meson approximation we can factorize the pionic
degrees of freedom and represent the amplitude Eq.(\ref{label2.3}) as
follows
\begin{eqnarray}\label{label2.4}
&&\frac{\displaystyle{\cal M}(\Lambda^+_c(Q) \to p(q)
K^-(q_-)\pi^+(q_+))}{\displaystyle \sqrt{2E_{\Lambda^+_c}V\,2E_p
V\,2E_{K^-} V\,2E_{\pi^+}V}} = <p(q) K^-(q_-)\pi^+(q_+)|{\cal L}_{\rm
eff}(0)|\Lambda^+_c(Q)>=\nonumber\\
&&= -\frac{G_F}{\sqrt{2}}\,V^*_{c s}\,V_{u
d}\,\bar{C}_1(\Lambda_{\chi})\,<p(q)
K^-(q_-)||\bar{s}(0)\,\gamma_{\mu}(1-\gamma^5)\,c(0)|\Lambda^+_c(Q)>\nonumber\\
&&\times <\pi^+(q_+)|\bar{u}(0)\,\gamma^{\mu}(1-\gamma^5)\,d(0)|0>,
\end{eqnarray}
where the matrix element
$<\pi^+(q_+)|\bar{u}(0)\,\gamma^{\mu}(1-\gamma^5)\,d(0)|0>$ can be
expressed in terms of the leptonic constant of the $\pi^+$--meson $F_{\pi}
=92.4\,{\rm MeV}$ [5]:
\begin{eqnarray}\label{label2.5}
\sqrt{2 E_{\pi^+}
V}<\pi^+(q_+)|\bar{u}(0)\,\gamma^{\mu}(1-\gamma^5)\,d(0)|0>=i\,\sqrt{2}\,F_{\pi}
\,q^{\mu}_+.
\end{eqnarray}
Substituting Eq.(\ref{label2.5}) in Eq.(\ref{label2.4}) we arrive at the
amplitude of the $\Lambda^+_c \to p + K^- + \pi^+$ decay
\begin{eqnarray}\label{label2.6}
&&\frac{\displaystyle{\cal M}(\Lambda^+_c(Q) \to p(q)
K^-(q_-)\pi^+(q_+))}{\displaystyle \sqrt{2E_{\Lambda^+_c}V\,2E_p
V\,2E_{K^-} V}} = \nonumber\\
&&=-i\,G_F\,V^*_{c s}\,V_{u
d}\,F_{\pi}\,\bar{C}_1(\Lambda_{\chi})\,q_+^{\mu}<p(q)
K^-(q_-)|\bar{s}(0)\,\gamma_{\mu}(1-\gamma^5)\,c(0)|\Lambda^+_c(Q)>,
\end{eqnarray}
where the matrix element $<p(q)
K^-(q_-)|\bar{s}(0)\,\gamma_{\mu}(1-\gamma^5)\,c(0)|\Lambda^+_c(Q)>$
describes the transition $\Lambda^+_c \to p + K^-$ induced by the current
$\bar{s}(0)\gamma^{\mu}(1-\gamma^5) c(0)$ and defined by strong low--energy
interactions.

\section{Strong low--energy transition $\Lambda^+_c \to p + K^-$}
\setcounter{equation}{0}

By applying the reduction technique we bring up the matrix element of the
strong low--energy transition $\Lambda^+_c \to p + K^-$  to the form
\begin{eqnarray}\label{label3.1}
\hspace{-0.5in}&&\sqrt{2E_{\Lambda^+_c}V\,2E_p V\,2E_{K^-}
V\,2E_{\pi^+}V}\,<p(q)
K^-(q_-)|\bar{s}(0)\,\gamma_{\mu}(1-\gamma^5)\,c(0)|\Lambda^+_c(Q)>=
\nonumber\\
\hspace{-0.5in}&&\lim_{\displaystyle Q^2\to M^2_{\Lambda^+_c}, q^2 \to
M^2_p, q^2_- \to M^2_K}i\int d^4x_1 d^4x_2 d^4x_3\,e^{\displaystyle iq\cdot
x_1}\,e^{\displaystyle  iq_- \cdot x_2}\,e^{\displaystyle  - i Q\cdot
x_3}\,\bar{u}_p(q,\sigma^{\prime}\,)\nonumber\\
\hspace{-0.5in}&&\overrightarrow{\Bigg(i\gamma^{\nu}\frac{\partial}{\partial x^{
\nu}_1} - M_p\Bigg)}(\Box_2 + M^2_K) <0|{\rm
T}(\psi_p(x_1)\varphi_{K^-}(x_2)[\bar{s}(0) \gamma^{\mu}(1-\gamma^5) c(0)]
\bar{\psi}_{\Lambda^+_c}(x_3))|0>\nonumber\\
\hspace{-0.5in}&&\overleftarrow{\Bigg(-
i\gamma^{\alpha}\frac{\partial}{\partial x^{\alpha}_3} -
M_{\Lambda^+_c}\Bigg)}u_{\Lambda^+_c}(Q,\sigma),
\end{eqnarray}
where $\psi_p(x_1)$, $\varphi_{K^-}(x_2)$ and
$\bar{\psi}_{\Lambda^+_c}(x_3)$ are the operators of the proton, the
$K^-$--meson and the $\Lambda^+_c$ interpolating fields,
$\bar{u}_p(q,\sigma^{\prime}\,)$ and $u_{\Lambda^+_c}(Q,\sigma)$ are the
Dirac bispinors of the proton and the $\Lambda^+_c$, respectively.

Following [6--11] in order to describe the r.h.s. of Eq.(\ref{label3.1}) at
the quark level we suggest to use the equations of motion
\begin{eqnarray}\label{label3.2}
\hspace{-0.5in}&&\overrightarrow{\Bigg(i\gamma^{\nu}\frac{\partial}{\partial x^{
\nu}_1} - M_p\Bigg)}\,\psi_p(x_1) = \frac{g_{\rm B}}{\sqrt{2}}\,\eta_{\rm
N}(x_1),\nonumber\\
\hspace{-0.5in}&&(\Box_2 + M^2_K)\,\varphi_{K^-}(x_2) =
\frac{g_{Kqq}}{\sqrt{2}}\,\bar{u}(x_2) i \gamma^5 s(x_2),\nonumber\\
\hspace{-0.5in}&&\bar{\psi}_{\Lambda^+_c}(x_3)\overleftarrow{\Bigg(-
i\gamma^{\alpha}\frac{\partial}{\partial x^{\alpha}_3} -
M_{\Lambda^+_c}\Bigg)}= \frac{g_{\rm
C}}{\sqrt{2}}\,\bar{\eta}_{\Lambda^+_c}(x_3).
\end{eqnarray}
Here $g_{\rm B}$ and $g_{\rm C}$ are the phenomenological coupling
constants of the proton and the $\Lambda^+_c$ coupled with three--quark
currents $\eta_{\rm N}(x_1) =
-\varepsilon^{ijk}[\bar{u^c}_i(x_1)\gamma^{\mu}u_j(x_1)]\gamma_{\mu}\gamma^5 d_k
(x_1)$ and $\bar{\eta}_{\Lambda^+_c}(x_3) =
\varepsilon^{ijk}\bar{c}_i(x_3)\gamma_{\mu}\gamma^5[\bar{d}_j(x_3)\gamma^{\mu}u^
c_k(x_3)]$ [11], respectively:
\begin{eqnarray}\label{label3.3}
\hspace{-0.5in}{\cal L}^{(\rm B)}_{\rm int}(x) = \frac{g_{\rm
B}}{\sqrt{2}}\,\bar{\psi}_p(x)\,\eta_{\rm N}(x) + \frac{g_{\rm
C}}{\sqrt{2}}\,\bar{\eta}_{\Lambda^+_c}(x)\,\psi_{\Lambda^+_c}(x) + {\rm
h.c.}.
\end{eqnarray}
Then, $i,j$ and $k$ are colour indices and $\bar{u^c}(x) = u(x)^T C$ and $C
= - C^T = - C^{\dagger} = - C^{-1} $ is a matrix of a charge conjugation,
$T$ is a transposition. The three--quark current $\eta_{\Lambda^+_c}(x) =
-\varepsilon^{ijk}[\bar{u^c}_i(x)\gamma^{\mu}d_j(x)]\gamma_{\mu}\gamma^5
c_k(x)$ coupled with the $\Lambda^+_c$ is constructed by analogy with
$\eta_{\rm N}(x) =
-\varepsilon^{ijk}[\bar{u^c}_i(x)\gamma^{\mu}u_j(x_1)]\gamma_{\mu}\gamma^5
d_k(x)$ due to the similar spinorial properties of the $\Lambda^+_c$ and
the proton. Then, $M_{\rm p} = 938\,{\rm MeV}$ and $M_{\Lambda^+_c} =
2285\,{\rm MeV}$ are the masses of the proton and the $\Lambda^+_c$.

The interaction of the $K^-$--meson with quarks is described by the
coupling constant $g_{Kqq}= \sqrt{2}m/F_{\pi}$
\begin{eqnarray}\label{label3.4}
\hspace{-0.5in}{\cal L}^{(\rm M)}_{\rm int}(x) =
\frac{g_{Kqq}}{\sqrt{2}}\,\bar{s}(x)\,i\gamma^5 u(x)\,\varphi_{K^-}(x)  +
\ldots + {\rm h.c.},
\end{eqnarray}
where $m = 330\,{\rm MeV}$ is the constituent quark mass calculated in the
chiral limit [5].

Substituting Eq.(\ref{label3.2}) in Eq.(\ref{label3.1}) we obtain
\begin{eqnarray}\label{label3.5}
\hspace{-0.5in}&&\sqrt{2E_{\Lambda^+_c}V\,2E_p V\,2E_{K^-}
V\,2E_{\pi^+}V}\,<p(q)
K^-(q_-)|\bar{s}(0)\,\gamma_{\mu}(1-\gamma^5)\,c(0)|\Lambda^+_c(Q)>=\nonumber\\
\hspace{-0.5in}&&=g_{\rm B}\,g_{\rm C}\frac{i}{2}\,\frac{m}{F_{\pi}}\int
d^4x_1 d^4x_2 d^4x_3\,e^{\displaystyle iq\cdot x_1}\,e^{\displaystyle  iq_-
\cdot x_2}\,e^{\displaystyle  - i Q\cdot x_3}\,\nonumber\\
\hspace{-0.5in}&&\bar{u}_p(q,\sigma^{\prime}\,)<0|{\rm T}(\eta_{\rm
N}(x_1)[\bar{u}(x_2)i\gamma^5 s(x_2)][\bar{s}(0) \gamma^{\mu}(1-\gamma^5)
c(0)] \bar{\eta}_{\Lambda^+_c}(x_3))|0>u_{\Lambda^+_c}(Q,\sigma),
\end{eqnarray}
where the external particles are kept on--mass shell, i.e. $Q^2 =
M^2_{\Lambda^+_c}$, $q^2 = M^2_p$ and $q^2_- = M^2_K$.

By applying the formulae of quark conversion [5] (Ivanov) we can determine
the vacuum expectation value in Eq.(\ref{label3.3}) in terms of the
constituent quark diagrams. In the momentum representation we get [5--11]:
\begin{eqnarray}\label{label3.6}
\hspace{-0.5in}&&\sqrt{2E_{\Lambda^+_c}V\,2E_p V\,2E_{K^-}
V\,2E_{\pi^+}V}\,<p(q)
K^-(q_-)|\bar{s}(0)\,\gamma_{\mu}(1-\gamma^5)\,c(0)|\Lambda^+_c(Q)>=\nonumber\\
\hspace{-0.5in}&&= - i g_{\rm B} g_{\rm C} \,\frac{3
m}{F_{\pi}}\Bigg[\frac{1}{16\pi^2}\Bigg]^2\int\frac{d^4k_1}{\pi^2i}\int
\frac{d^4k_2}{\pi^2i}\bar{u}_p(q,\sigma^{\prime}\,) \gamma_{\alpha}\gamma^5
\frac{1}{m - \hat{k}_1}\gamma^{\beta}\frac{1}{m +
\hat{k}_2}\gamma^{\alpha}\frac{1}{m - \hat{q} + \hat{k}_1 +
\hat{k}_2}\nonumber\\
\hspace{-0.5in}&&\gamma^5 \frac{1}{m - \hat{q} - \hat{q}_- + \hat{k}_1 +
\hat{k}_2}\gamma_{\mu}(1 - \gamma^5) \frac{1}{M_c - \hat{Q} + \hat{k}_1 +
\hat{k}_2} \gamma_{\beta}\gamma^5\,u_{\Lambda^+_c}(Q,\sigma),
\end{eqnarray}
where $M_c = 1860\,{\rm MeV}$ is the mass of the constituent $c$--quark
[7--11]. In the HQET the r.h.s. of Eq.(\ref{label2.4}) is given by
[3,4,7--11]:
\begin{eqnarray}\label{label3.7}
\hspace{-0.5in}&&\sqrt{2E_{\Lambda^+_c}V\,2E_p V\,2E_{K^-}
V\,2E_{\pi^+}V}\,<p(q)
K^-(q_-)|\bar{s}(0)\,\gamma_{\mu}(1-\gamma^5)\,c(0)|\Lambda^+_c(Q)>=
\nonumber\\
\hspace{-0.5in}&&= - i g_{\rm B} g_{\rm C}\,\frac{3
m}{F_{\pi}}\Bigg[\frac{1}{16\pi^2}\Bigg]^2\int\frac{d^4k_1}{\pi^2i}\int
\frac{d^4k_2}{\pi^2i}\,\bar{u}_p(q,\sigma^{\prime}\,)
\gamma_{\alpha}\gamma^5 \frac{1}{m - \hat{k}_1}\gamma^{\beta}\frac{1}{m +
\hat{k}_2}\gamma^{\alpha}\frac{1}{m - \hat{q} + \hat{k}_1 +
\hat{k}_2}\nonumber\\
\hspace{-0.5in}&&\gamma^5 \frac{1}{m - \hat{q} - \hat{q}_- + \hat{k}_1 +
\hat{k}_2}\gamma_{\mu}(1 - \gamma^5) \Bigg(\frac{1+
\hat{v}}{2}\Bigg)\frac{1}{[(k_1 + k_2)\cdot v + i0]} \gamma_{\beta}\gamma^5
\,u_{\Lambda^+_c}(Q,\sigma),
\end{eqnarray}
where $v^{\mu}$ is 4--velocity of the $\Lambda^+_c$ [3,4,7--11] normalized
by $v^{\mu}v_{\mu} = 1$.

For the computation of the momentum integral we assume [12] that the proton
is a very heavy state and its 4--momentum is much larger than other momenta
in the integrand of Eq.(\ref{label3.7}). Keeping the leading terms in the
large $M_p$ expansion we reduce the r.h.s. of Eq.(\ref{label3.7}) to the
form
\begin{eqnarray}\label{label3.8}
\hspace{-0.5in}&&\sqrt{2E_{\Lambda^+_c}V\,2E_p V\,2E_{K^-}
V\,2E_{\pi^+}V}\,<p(q)
K^-(q_-)|\bar{s}(0)\,\gamma_{\mu}(1-\gamma^5)\,c(0)|\Lambda^+_c(Q)>=
\nonumber\\
\hspace{-0.5in}&&= i \frac{g_{\rm B} g_{\rm C}}{M^2_p} \,\frac{3
m}{F_{\pi}}\Bigg[\frac{1}{16\pi^2}\Bigg]^2\int\frac{d^4k_1}{\pi^2i}\int
\frac{d^4k_2}{\pi^2i}\,\bar{u}_p(q,\sigma^{\prime}\,)
\gamma_{\alpha}\gamma^5 \frac{1}{m - \hat{k}_1}\gamma^{\beta}\frac{1}{m +
\hat{k}_2}\gamma^{\alpha}\gamma_{\mu}(1 - \gamma^5) \nonumber\\
\hspace{-0.5in}&&\Bigg(\frac{1+ \hat{v}}{2}\Bigg)\frac{1}{[(k_1 + k_2)\cdot
v + i0]} \gamma_{\beta}\gamma^5 \,u_{\Lambda^+_c}(Q,\sigma).
\end{eqnarray}
The replacement of the constituent quark Green function
\begin{eqnarray}\label{label3.9}
\frac{1}{m - \hat{q} + \hat{k}_1 + \hat{k}_2} \to -
\frac{\hat{p}_1}{M^2_{\rm p}}
\end{eqnarray}
agrees with the heavy baryon [16,17] and HQET [3,4] approaches. Indeed, in
accordance with [16,17] and HQET [3,4] we obtain
\begin{eqnarray}\label{label3.10}
&&\frac{1}{m - \hat{q} + \hat{k}_1 + \hat{k}_2} = \frac{m + \hat{q} -
\hat{k}_1 - \hat{k}_2}{m^2 - (q - k_1 - k_2)^2 - i0} = \nonumber\\
&&= - \frac{m + \hat{q} - \hat{k}_1 - \hat{k}_2}{M^2_{\rm p} - 2(k_1 +
k_2)\cdot q - m^2 + (k_1 + k_2)^2 + i0}=\nonumber\\
&&=- \frac{1}{M_{\rm p}}\,\frac{\displaystyle \frac{m - \hat{k}_1 -
\hat{k}_2}{M_{\rm p}} + \hat{v}}{\displaystyle 1 + \frac{2(k_1 + k_2)\cdot
v}{M_{\rm p}} - \frac{m^2}{M^2_{\rm p}} + \frac{(k_1 + k_2)^2}{M^2_{\rm p}}
+ i0},
\end{eqnarray}
where we have set $q^{\mu} = M_{\rm p}\,v^{\mu}$ [3,4,16,17]. In the case
of $m = M_{\rm p}$ and in the limit  $M_{\rm p} \to \infty$ we arrive at
the well--known expression for the Green function of a heavy baryon (or a
heavy quark in HQET [3,4]) [16,17]:
\begin{eqnarray}\label{label3.11}
\frac{1}{m - \hat{q} + \hat{k}_1 + \hat{k}_2} &=&\frac{1}{M_{\rm p} +
\hat{k}_1 + \hat{k}_2 - M_{\rm p}\hat{v}} \to \nonumber\\
&\to& \Bigg(\frac{1 + \hat{v}}{2}\Bigg)\,\frac{1}{(k_1 + k_2)\cdot v + i0}.
\end{eqnarray}
In our case $m\ll M_{\rm p}$, therefore, in the limit $M_{\rm p} \to
\infty$ [16,17] we arrive at the expression Eq.(\ref{label3.9}).

The computation of the momentum integrals in Eq.(\ref{label3.8}) is carried
out in the Appendix. Thus, the matrix element $<p(q)
K^-(q_-)|\bar{s}(0)\,\gamma_{\mu}(1-\gamma^5)\,c(0)|\Lambda^+_c(Q)>$ is
defined
\begin{eqnarray}\label{label3.12}
\hspace{-0.5in}&&\sqrt{2E_{\Lambda^+_c}V\,2E_p V\,2E_{K^-}
V\,2E_{\pi^+}V}\,<p(q)
K^-(q_-)|\bar{s}(0)\,\gamma_{\mu}(1-\gamma^5)\,c(0)|\Lambda^+_c(Q)>=
\nonumber\\
\hspace{-0.5in}&&= i g_{\rm \pi NN}\,\frac{2}{5}\,\frac{g_{\rm C}}{g_{\rm
B}}\, \frac{\Lambda_{\chi}}{m^2}\,\bar{u}_p(q,\sigma^{\prime}\,)
\gamma_{\alpha}\gamma^5
\hat{v}\gamma^{\beta}\hat{v}\gamma^{\alpha}\gamma_{\mu}(1 - \gamma^5)
\Bigg(\frac{1+ \hat{v}}{2}\Bigg)\gamma_{\beta}\gamma^5
\,u_{\Lambda^+_c}(Q,\sigma),
\end{eqnarray}
where $g_{\rm \pi NN}$ is the coupling constant of the ${\rm \pi NN}$
interaction expressed in terms of the parameters of the model through the
relation $g_{\rm \pi NN} = g^2_{\rm B}(2m/3F_{\pi})(<\bar{q}q>^2/M^2_p)$
[12]. After some algebra the matrix element Eq.(\ref{label3.12}) amounts to
\begin{eqnarray}\label{label3.13}
\hspace{-0.5in}&&\sqrt{2E_{\Lambda^+_c}V\,2E_p V\,2E_{K^-}
V\,2E_{\pi^+}V}\,<p(q)
K^-(q_-)|\bar{s}(0)\,\gamma_{\mu}(1-\gamma^5)\,c(0)|\Lambda^+_c(Q)>=
\nonumber\\
\hspace{-0.5in}&&= i g_{\rm \pi NN}\,\frac{4}{5}\,\frac{g_{\rm C}}{g_{\rm
B}}\,\frac{\Lambda_{\chi}}{m^2}\,\bar{u}_p(q,\sigma^{\prime}\,)
\,[2\,v_{\mu}(1 - \gamma^5) + \gamma_{\mu}(1 +
\gamma^5)]\,u_{\Lambda^+_c}(Q,\sigma) =\nonumber\\
\hspace{-0.5in}&&= i g_{\rm \pi NN}\,\frac{4}{5}\,\frac{g_{\rm C}}{g_{\rm
B}}\,\frac{\Lambda_{\chi}}{m^2}\,\bar{u}_p(q,\sigma^{\prime}\,) \,(1 -
\gamma^5)\,(2\,v_{\mu} + \gamma_{\mu})\,u_{\Lambda^+_c}(Q,\sigma),
\end{eqnarray}
where we used the Dirac equation of motion $\hat{v}
u_{\Lambda^+_c}(Q,\sigma) = u_{\Lambda^+_c}(Q,\sigma)$ for the free
$\Lambda^+_c$.

\section{Proton energy spectrum of the $\Lambda^+_c \to p + K^- + \pi^+$ decay}
\setcounter{equation}{0}

The amplitude of the $\Lambda^+_c \to p + K^- + \pi^+$ decay is given by
\begin{eqnarray}\label{label4.1}
&&{\cal M}(\Lambda^+_c(Q) \to p(q) K^-(q_-)\pi^+(q_+)) =  G_F\,V^*_{c
s}\,V_{u d}\,\bar{C}_1(\Lambda_{\chi})\,\nonumber\\
&&\times \,\frac{4}{5}\,\frac{g_{\rm \pi
NN}}{M_{\Lambda^+_c}}\,\Bigg[\frac{g_{\rm C}}{g_{\rm
B}}\,\frac{F_{\pi}\Lambda_{\chi}}{m^2}\Bigg]\,\bar{u}_p(q,\sigma^{\prime}\,) \,(
1 - \gamma^5)\,(2 Q\cdot q_+ +
M_{\Lambda^+_c}\hat{q}_+)\,u_{\Lambda^+_c}(Q,\sigma).
\end{eqnarray}
The partial width of the $\Lambda^+_c \to p + K^- + \pi^+$ decay determined
in the rest frame of the $\Lambda^+_c$ reads
\begin{eqnarray}\label{label4.2}
&&d\Gamma(\Lambda^+_c \to p\,K^- \pi^+) = \frac{1}{2M_{\Lambda^+_c}}
\overline{|{\cal M}(\Lambda^+_c(Q) \to p(q) K^-(q_-)\pi^+(q_+))|^2} \nonumber\\
&&(2\pi)^4\,\delta^{(4)}(Q - q - q_- - q_+)\,\frac{d^3q}{(2\pi)^3
2E_p}\,\frac{d^3q_-}{(2\pi)^3 2E_{K^-}}\,\frac{d^3q_+}{(2\pi)^3 2E_{\pi^+}},
\end{eqnarray}
We define the quantity $\overline{|{\cal M}(\Lambda^+_c(Q) \to p(q)
K^-(q_-)\pi^+(q_+))|^2}$ for the polarized $\Lambda^+_c$ and the
unpolarized proton:
\begin{eqnarray}\label{label4.3}
&&\overline{|{\cal M}(\Lambda^+_c(Q) \to p(q) K^-(q_-)\pi^+(q_+))|^2} =
|G_F\,V^*_{c s}\,V_{u d}\,\bar{C}_1(\Lambda_{\chi})|^2
\Bigg[\frac{4}{5}\,\frac{g_{\rm \pi NN}}{M_{\Lambda^+_c}}\,\frac{g_{\rm
C}}{g_{\rm B}}\,\frac{F_{\pi}\Lambda_{\chi}}{m^2}\Bigg]^2\,\nonumber\\
&&\times \,\frac{1}{2}\,{\rm tr}\{(\hat{Q} + M_{\Lambda^+_c})(1 +
\gamma^5\hat{\omega}_{\Lambda^+_c})(2 Q\cdot q_+ +
M_{\Lambda^+_c}\hat{q}_+)(1 + \gamma^5)(\hat{q} + M_p) (1 -
\gamma^5)\nonumber\\
&&\times \,(2 Q\cdot q_+ + M_{\Lambda^+_c}\hat{q}_+)\},
\end{eqnarray}
where $\omega^{\mu}_{\Lambda^+_c}$ is the space--like unit 4--vector
$(\omega^2_{\Lambda^+_c}= - 1)$ orthogonal to the 4--momentum of the
$\Lambda^+_c$ $(\omega_{\Lambda^+_c}\cdot Q = 0)$ and related to the
direction of its spin. In the rest frame of the $\Lambda^+_c$ we have:
$\omega^{\mu}_{\Lambda^+_c} = (0, \vec{\omega}_{\Lambda^+_c})$ such as
$\vec{\omega}^{\,2}_{\Lambda^+_c} = 1$.

Neglecting the terms proportional to $M^2_{\pi}$ and $M^2_K$ the result of
the calculation of the trace reads
\begin{eqnarray}\label{label4.4}
\frac{1}{2}\,{\rm tr}\{\ldots\} &=& 16(Q\cdot q_+)^2(Q\cdot q) + 24
M^2_{\Lambda^+_c}(Q\cdot q_+)(q\cdot q_+) - 32 M_{\Lambda^+_c}(Q\cdot
q_+)^2 (\omega_{\Lambda^+_c}\cdot q) \nonumber\\
&+& 16 M_{\Lambda^+_c} (Q\cdot q_+)(Q\cdot q)(\omega_{\Lambda^+_c} \cdot
q_+) + 8 M^3_{\Lambda^+_c}(q\cdot q_+) (\omega_{\Lambda^+_c} \cdot q_+).
\end{eqnarray}
For the derivation of the proton energy spectrum it is convenient to use
the formula
\begin{eqnarray}\label{label4.5}
\hspace{-0.5in}&&\int q^{\alpha}_+ q^{\beta}_+\,\delta^{(4)}(Q - q - q_- -
q_+) \frac{d^3q_-}{2E_{K^-}}\frac{d^3q_+}{2E_{\pi^+}} =\nonumber\\
\hspace{-0.5in}&&= \frac{1}{12}\,(- (Q-q)^2\,g^{\alpha\beta} +
4\,(Q-q)^{\alpha}(Q-q)^{\beta})\int \delta^{(4)}(Q - q - q_- - q_+)
\frac{d^3q_-}{2E_{K^-}}\frac{d^3q_+}{2E_{\pi^+}}=\nonumber\\
\hspace{-0.5in}&&= \frac{\pi}{24}\,\times\,\Big[- (Q-q)^2\,g^{\alpha\beta}
+ 4\,(Q-q)^{\alpha}(Q-q)^{\beta}\Big],
\end{eqnarray}
which is valid when the contributions proportional to $M^2_{\pi}$ and
$M^2_K$ are neglected. Using Eq.(\ref{label4.5}) we get
\begin{eqnarray}\label{label4.6}
\hspace{-0.5in}&&\int \frac{1}{2}\,{\rm tr}\{\ldots\}\,\delta^{(4)}(Q - q -
q_- - q_+) \frac{d^3q_-}{2E_{K^-}}\frac{d^3q_+}{2E_{\pi^+}} =\nonumber\\
\hspace{-0.5in}&&= \frac{\pi}{3}\,\Big[(15 M^4_{\Lambda^+_c}(Q\cdot q) - 18
M^2_{\Lambda^+_c}(Q\cdot q)^2 + 8 (Q\cdot q)^3 + 7 M^2_{\Lambda^+_c} M^2_p
(Q\cdot q) -12 M^4_{\Lambda^+_c} M^2_p)\nonumber\\
\hspace{-0.5in}&& - (\omega_{\Lambda^+_c}\cdot q) M_{\Lambda^+_c}( 13
M^4_{\Lambda^+_c} - 14 M^2_{\Lambda^+_c}(Q\cdot q) + 8 (Q\cdot q)^3 - 7
M^2_{\Lambda^+_c} M^2_p)\Big] =\nonumber\\
\hspace{-0.5in}&& = \frac{5\pi}{2} M^6_{\Lambda^+_c} x\,\Bigg[\Bigg(1 -
\frac{3}{5}\,x + \frac{2}{15}\,x^2 + \frac{7}{60}\,\xi^2 -
\frac{2}{5}\frac{\xi^2}{x}\Bigg)\nonumber\\
\hspace{-0.5in}&&+ (\vec{\omega}_{\Lambda^+_c}\cdot \vec{n}_p)
\frac{13}{15}{\displaystyle \sqrt{1 - \frac{\xi^2}{x^2}}}\Bigg(1 -
\frac{7}{13}\,x + \frac{2}{13}\,x^2  - \frac{7}{52}\,\xi^2\Bigg)\Bigg],
\end{eqnarray}
where the final expression is taken in the rest frame of the $\Lambda^+_c$,
$x = 2 E_p/M_{\Lambda^+_c}$ is the scaled proton energy, $\xi = 2
M_p/M_{\Lambda^+_c}$ and $\vec{n}_p = \vec{q}/|\vec{q}\,|$. The scaled
proton energy $x$ ranges the region $\xi \le x\le 1 + \xi^2/4$.

The proton energy spectrum of the $\Lambda^+_c \to p + K^- + \pi^+$ decay
in the rest frame of the $\Lambda^+_c$ is determined:
\begin{eqnarray}\label{label4.7}
\hspace{-0.5in}&&d\Gamma(\Lambda^+_c \to p\,K^- \pi^+) = |G_F\,V^*_{c
s}\,V_{u d}\,\bar{C}_1(\Lambda_{\chi})|^2 \Bigg[g_{\rm \pi
NN}\,\frac{4}{5}\,\frac{g_{\rm C}}{g_{\rm
B}}\,\frac{F_{\pi}\Lambda_{\chi}}{m^2}\Bigg]^2\,\times\,\Bigg[\frac{5
M^5_{\Lambda^+_c}}{512\pi^3}\Bigg]\,\nonumber\\
\hspace{-0.5in}&&\times\,\Bigg(1 - \frac{3}{5}\,x + \frac{2}{15}\,x^2 +
\frac{7}{60}\,\xi^2 - \frac{2}{5}\,\frac{\xi^2}{x}\Bigg)\,\Big[1 +
\alpha(x)\,(\vec{\omega}_{\Lambda^+_c}\cdot \vec{n}_p)\Big]\,x\,\sqrt{x^2 -
\xi^2}\,d x\,\frac{d\Omega_{\vec{n}_p}}{4\pi},
\end{eqnarray}
where $d\Omega_{\vec{n}_p}$ is the solid angle of the  unit vector
$\vec{n}_p = \vec{q}/|\vec{q}\,|$ and $\alpha(x)$, the parameter of the
asymmetry related to the polarization of the $\Lambda^+_c$, is given by
\begin{eqnarray}\label{label4.8}
\hspace{-0.5in}\alpha(x) = \frac{13}{15}{\displaystyle \sqrt{1 -
\frac{\xi^2}{x^2}}}\frac{\displaystyle 1 - \frac{7}{13}\,x +
\frac{2}{13}\,x^2  - \frac{7}{52}\,\xi^2}{\displaystyle 1 - \frac{3}{5}\,x
+ \frac{2}{15}\,x^2 + \frac{7}{60}\,\xi^2 - \frac{2}{5}\frac{\xi^2}{x}}.
\end{eqnarray}
In order to apply Eq.(\ref{label4.7}) to the analysis of the polarization
of the $\Lambda^+_c$ in the processes of photo and hadroproduction we have
to remove an uncertainty related to the arbitrary coupling constant
$g^2_{\rm C}$. For this aim we suggest to normalize the proton energy
spectrum to the partial width of the mode $\Lambda^+_c \to p + K^- + \pi^+$
and replace the coupling constant $g^2_{\rm C}$ by the experimental value
of the probability. Integrating Eq.(\ref{label4.7}) over all variables we
obtain the partial width of the mode:
\begin{eqnarray}\label{label4.9}
\hspace{-0.3in}\Gamma(\Lambda^+_c \to p\,K^- \pi^+) = |G_F\,V^*_{c s}\,V_{u
d}\,\bar{C}_1(\Lambda_{\chi})|^2 \,\Bigg[g_{\rm \pi
NN}\,\frac{4}{5}\,\frac{g_{\rm C}}{g_{\rm
B}}\,\frac{F_{\pi}\Lambda_{\chi}}{m^2}\Bigg]^2\,\times\,\Bigg[\frac{5
M^5_{\Lambda^+_c}}{512\pi^3}\Bigg]\,\times\,f(\xi).
\end{eqnarray}
The function $f(\xi)$ is determined by the integral
\begin{eqnarray}\label{label4.10}
\hspace{-0.5in}f(\xi) = \int\limits^{1 + \xi^2/4}_{\xi}\Bigg(1 -
\frac{3}{5}\,x + \frac{2}{15}\,x^2 + \frac{7}{60}\,\xi^2 -
\frac{2}{5}\,\frac{\xi^2}{x}\Bigg)\,x\,\sqrt{x^2 - \xi^2}\,dx = 0.065.
\end{eqnarray}
The numerical value has been obtained at $M_{\Lambda^+_c}=2285\,{\rm MeV}$
and $M_p = 938\,{\rm MeV}$.

The proton energy spectrum of the  $\Lambda^+_c \to p + K^- + \pi^+$ decay
is then given by
\begin{eqnarray}\label{label4.11}
\hspace{-0.5in}\frac{dB(\Lambda^+_c \to p\,K^- \pi^+)}{B(\Lambda^+_c \to
p\,K^- \pi^+)} &=& 15.40\,\times\,\Bigg(1 - \frac{3}{5}\,x +
\frac{2}{15}\,x^2 + \frac{7}{60}\,\xi^2 -
\frac{2}{5}\,\frac{\xi^2}{x}\Bigg) \nonumber\\
\hspace{-0.5in}&&\times\,\Big[1 +
\alpha(x)\,(\vec{\omega}_{\Lambda^+_c}\cdot \vec{n}_p)\Big]\,x\,\sqrt{x^2 -
\xi^2}\,d x\,\frac{d\Omega_{\vec{n}_p}}{4\pi}.
\end{eqnarray}
By using the experimental value of the probability $B(\Lambda^+_c \to
p\,K^- \pi^+)_{\exp}=(0.050\pm 0.013)$ [2] we obtain
\begin{eqnarray}\label{label4.12}
\hspace{-0.3in}dB(\Lambda^+_c \to p\,K^- \pi^+) &=&(0.77\pm
0.20)\,\times\,\Bigg(1 - \frac{3}{5}\,x + \frac{2}{15}\,x^2 +
\frac{7}{60}\,\xi^2 - \frac{2}{5}\,\frac{\xi^2}{x}\Bigg) \nonumber\\
\hspace{-0.5in}&&\times\,\Big[1 +
\alpha(x)\,(\vec{\omega}_{\Lambda^+_c}\cdot \vec{n}_p)\Big]\,x\,\sqrt{x^2 -
\xi^2}\,d x\,\frac{d\Omega_{\vec{n}_p}}{4\pi}.
\end{eqnarray}
Integrating over the energy of the proton we derive the angular
distribution of the probability of the $\Lambda^+_c \to p + K^- + \pi^+$
decay:
\begin{eqnarray}\label{label4.13}
\hspace{-0.5in}\frac{dB(\Lambda^+_c \to p\,K^- \pi^+)}{d\Omega_{\vec{n}_p}}
= \frac{0.050\pm 0.013}{4\pi}\,\times\,\Big[1 +
0.77\,(\vec{\omega}_{\Lambda^+_c}\cdot \vec{n}_p)\Big].
\end{eqnarray}
Thus, the proton energy spectrum  Eq.(\ref{label4.12}) and the angular
distribution of the probability Eq.(\ref{label4.13}) of the $\Lambda^+_c
\to p + K^- + \pi^+$ decay do not contain arbitrary parameters and,
therefore, can be applied to the analysis of the polarization of the
$\Lambda^+_c$ in the processes of photo and hadroproduction.

The formulae Eq.(\ref{label4.12}) and Eq.(\ref{label4.13}) define the
polarization of the $\Lambda^+_c$ relative to the momentum of the proton.
If the spin of the $\Lambda^+_c$ is parallel to the momentum of the proton,
the right--handed (R) polarization, the scalar product
$\vec{\omega}_{\Lambda^+_c}\cdot \vec{n}_p$ amounts to
$\vec{\omega}_{\Lambda^+_c}\cdot \vec{n}_p = \cos\vartheta$. The angular
distribution  of the probability  reads
\begin{eqnarray}\label{label4.14}
\hspace{-0.5in}\frac{dB(\Lambda^+_c \to p\,K^- \pi^+)_{\rm
(R)}}{d\Omega_{\vec{n}_p}} = \frac{0.050\pm 0.013}{4\pi}\,\times\,(1 +
0.77\,\cos\theta).
\end{eqnarray}
In turn, for the left--handed (L) polarization of the $\Lambda^+_c$, the
spin of the $\Lambda^+_c$ is anti--parallel to the momentum of the proton,
the scalar product reads $(\vec{\omega}_{\Lambda^+_c}\cdot \vec{n}_p) =
-\cos\vartheta$ and the angular distribution becomes equal
\begin{eqnarray}\label{label4.15}
\hspace{-0.5in}\frac{dB(\Lambda^+_c \to p\,K^- \pi^+)_{\rm
(L)}}{d\Omega_{\vec{n}_p}} = \frac{0.050\pm 0.013}{4\pi}\,\times\,(1 -
0.77\,\cos\theta).
\end{eqnarray}
For the right-- and left--handed polarizations of the $\Lambda^+_c$ the
proton energy spectrum is given by
\begin{eqnarray}\label{label4.16}
\hspace{-0.3in}dB(\Lambda^+_c \to p\,K^- \pi^+)_{\rm (R,L)} &=&(0.77\pm
0.20)\,\times\,\Bigg(1 - \frac{3}{5}\,x + \frac{2}{15}\,x^2 +
\frac{7}{60}\,\xi^2 - \frac{2}{5}\,\frac{\xi^2}{x}\Bigg) \nonumber\\
\hspace{-0.5in}&&\times\,\Big[1 \pm
\alpha(x)\,\cos\vartheta\Big]\,x\,\sqrt{x^2 - \xi^2}\,d
x\,\frac{d\Omega_{\vec{n}_p}}{4\pi}.
\end{eqnarray}
The formulae Eq.(\ref{label4.12}) and Eq.(\ref{label4.13}) resemble the
electron energy spectrum and the angular distribution of the probability of
the $\beta$--decay of the $\mu^-$--meson, i.e. $\mu^- \to {\rm e}^- +
\bar{\nu}_{\rm e} + \nu_{\mu}$. Therefore, the procedure of the restoration
of the polarization of the $\Lambda^+_c$ in the $\Lambda^+_c \to p + K^-
\pi^+$ decay is in complete analogy to the procedure of the measurement of
the polarization of the $\mu^-$--meson in the $\mu^- \to {\rm e}^- +
\bar{\nu}_{\rm e} + \nu_{\mu}$ decay [2].

\section{Probability of the $\Lambda^+_c \to p + \bar{K}^0$ decay}
\setcounter{equation}{0}

Most modes of the  $\Lambda^+_c$ decays are measured relative to the mode
$\Lambda^+_c \to p + K^- + \pi^+$ [2]. For the theoretical description of
the $\Lambda^+_c \to p + K^- + \pi^+$ decay at the quark level we have
introduced the phenomenological low--energy interaction of the
$\Lambda^+_c$ with the three--quark current $\eta_{\Lambda^+_c}(x)$
containing an arbitrary phenomenological coupling constant $g_{\rm C}$
Eq.(\ref{label3.3}). The spinorial structure of the three--quark current
${\eta}_{\Lambda^+_c}(x) = -
\varepsilon^{ijk}[\bar{u^c}_i(x)\gamma^{\mu}d_j(x)]\gamma_{\mu}\gamma^5
c_k(x)$ defined as the product of the axial--vector light diquark density
$[\bar{u^c}_i(x) \gamma^{\mu} d_j(x)]$ transforming under $SU(3)_f\times
SU(3)_c$ group like $(\underline{6}_f,\tilde{\underline{3}}_c)$ and the
$c$--quark field $c_k(x)$ is caused by the dynamics of the quark
confinement given by a linearly rising interquark potential [6,11]. In
order to verify the validity of the approach applied to the computation of
the proton energy spectrum and the angular distribution of the probability
of the mode $\Lambda^+_c \to p + K^- + \pi^+$ it is not sufficient to be
restricted by the consideration only this mode. For the confirmation of the
result obtained for the mode $\Lambda^+_c \to p + K^- + \pi^+$ one needs
the computation of the probabilities of other modes relative to the
probability of the main mode $\Lambda^+_c \to p + K^- + \pi^+$. In the
ratio the coupling constant $g_{\rm C}$ cancels itself and the theoretical
result turns out to be dependent on the Wilson coefficients, determined by
the short--distance quark--gluon interactions, and the long--distance
dynamics, described by the effective quark model with chiral $U(3)\times
U(3)$ symmetry motivated by QCD with a linearly rising confinement
potential. The agreement between the experimental data and the theoretical
predictions for the ratios should testify both the self--consistency of the
approach and the consistency of it with a short--distance QCD. Below we
obtain the evidence of the self--consistency and the consistency of the
approach by example of the calculation of the decay mode $\Lambda^+_c \to p
+ \bar{K}^0$. We have chosen this mode due to the following reasons. First,
it contains the proton in the final state, and, second, the computation of
the matrix element of this mode is completely different to the computation
of the matrix element of the mode $\Lambda^+_c \to p + K^- + \pi^+$.
Indeed, the mode $\Lambda^+_c \to p + \bar{K}^0$ unlike the mode
$\Lambda^+_c \to p + K^- + \pi^+$ admits the factorization of the baryonic
and mesonic degrees of freedom.

The effective low--energy Lagrangian responsible for the decay $\Lambda^+_c
\to p + \bar{K}^0$ can be obtained  from the effective Lagrangian
Eq.(\ref{label2.1}) at $\mu = \Lambda_{\chi}$:
\begin{eqnarray}\label{label5.1}
{\cal L}_{\rm eff}(x) = - \frac{G_F}{\sqrt{2}}\,V^*_{c s}\,V_{u
d}\,\bar{C}_2(\Lambda_{\chi})\,[\bar{u}(x)\,\gamma_{\mu}(1-\gamma^5)\,c(x)]
\,[\bar{s}(x) \gamma^{\mu}(1-\gamma^5)\,d(x)],
\end{eqnarray}
where $\bar{C}_2(\Lambda_{\chi}) = C_2(\Lambda_{\chi}) +
C_1(\Lambda_{\chi})/3$.

The amplitude of the $\Lambda^+_c \to p + \bar{K}^0$ decay can be defined
by analogy with Eq.(\ref{label2.6}):
\begin{eqnarray}\label{label5.2}
&&\frac{\displaystyle{\cal M}(\Lambda^+_c(Q) \to p(q)
\bar{K}^0(q_0))}{\displaystyle \sqrt{2E_{\Lambda^+_c}V\,2E_p V}} =
\nonumber\\
&&=-i\,G_F\,V^*_{c s}\,V_{u
d}\,\bar{C}_2(\Lambda_{\chi})\,F_K\,q_0^{\mu}<p(q)
|\bar{u}(0)\,\gamma_{\mu}(1-\gamma^5)\,c(0)|\Lambda^+_c(Q)>,
\end{eqnarray}
where $F_K = 113\,{\rm MeV}$ [2] is the leptonic constant of the $K$--mesons.

To the computation of the matrix element $<p(q)
|\bar{u}(0)\,\gamma_{\mu}(1-\gamma^5)\,c(0)|\Lambda^+_c(Q)>$ we apply the
reduction technique. By using the equations of motion Eq.(\ref{label2.2})
we arrive at the expression
\begin{eqnarray}\label{label5.3}
\hspace{-0.5in}&&\sqrt{2E_{\Lambda^+_c}V\,2E_p V}\,<p(q)
|\bar{u}(0)\,\gamma_{\mu}(1-\gamma^5)\,c(0)|\Lambda^+_c(Q)> = \nonumber\\
\hspace{-0.5in}&&= g_{\rm B}\,g_{\rm C}\frac{1}{2}\,\int d^4x_1 d^4x_2
\,e^{\displaystyle  iq\cdot x_1}\,e^{\displaystyle  - i Q\cdot
x_2}\nonumber\\
\hspace{-0.5in}&&\times\,\bar{u}_p(q,\sigma^{\prime}\,) <0|{\rm
T}(\eta_{\rm N}(x_1)[\bar{u}(0) \gamma^{\mu}(1-\gamma^5) c(0)]
\bar{\eta}_{\Lambda^+_c}(x_2))|0>u_{\Lambda^+_c}(Q,\sigma),
\end{eqnarray}
where the $\Lambda^+_c$ and the proton are kept on--mass shell, i.e. $Q^2 =
M^2_{\Lambda^+_c}$ and $q^2 = M^2_p$. In terms of the constituent quark
diagrams represented by the momentum integrals the r.h.s. of
Eq.(\ref{label5.3}) defined in HQET reads
\begin{eqnarray}\label{label5.4}
\hspace{-0.5in}&&\sqrt{2E_{\Lambda^+_c}V\,2E_p V}\,<p(q)
|\bar{u}(0)\,\gamma_{\mu}(1-\gamma^5)\,c(0)|\Lambda^+_c(Q)>=  \nonumber\\
\hspace{-0.5in}&&= - 3\,g_{\rm B} g_{\rm
C}\,\Bigg[\frac{1}{16\pi^2}\Bigg]^2\int\frac{d^4k_1}{\pi^2i}\int
\frac{d^4k_2}{\pi^2i}\,\bar{u}_p(q,\sigma^{\prime}\,)
\gamma_{\alpha}\gamma^5 \frac{1}{m - \hat{k}_1}\gamma^{\beta}\frac{1}{m +
\hat{k}_2}\gamma^{\alpha}\frac{1}{m - \hat{q} + \hat{k}_1 +
\hat{k}_2}\nonumber\\
\hspace{-0.5in}&&\gamma_{\mu}(1 - \gamma^5) \Bigg(\frac{1+
\hat{v}}{2}\Bigg)\frac{1}{[(k_1 + k_2)\cdot v + i0]} \gamma_{\beta}\gamma^5
\,u_{\Lambda^+_c}(Q,\sigma).
\end{eqnarray}
Keeping the leading terms in the large $M_p$ expansion we reduce the r.h.s.
of Eq.(\ref{label5.4}) to the form
\begin{eqnarray}\label{label5.5}
\hspace{-0.5in}&&\sqrt{2E_{\Lambda^+_c}V\,2E_p V}\,<p(q)
|\bar{u}(0)\,\gamma_{\mu}(1-\gamma^5)\,c(0)|\Lambda^+_c(Q)>= \nonumber\\
\hspace{-0.5in}&&= 3\,\frac{g_{\rm B} g_{\rm
C}}{M^2_p}\,\Bigg[\frac{1}{16\pi^2}\Bigg]^2\int\frac{d^4k_1}{\pi^2i}\int
\frac{d^4k_2}{\pi^2i}\,\bar{u}_p(q,\sigma^{\prime}\,)
\gamma_{\alpha}\gamma^5 \frac{1}{m - \hat{k}_1}\gamma^{\beta}\frac{1}{m +
\hat{k}_2}\gamma^{\alpha}\,\hat{q}\gamma_{\mu}(1 - \gamma^5)\nonumber\\
\hspace{-0.5in}&&\times\,\Bigg(\frac{1+ \hat{v}}{2}\Bigg)\frac{1}{[(k_1 +
k_2)\cdot v + i0]} \gamma_{\beta}\gamma^5 \,u_{\Lambda^+_c}(Q,\sigma).
\end{eqnarray}
The integrals over $k_1$ and $k_2$ have been calculated in the Appendix.
Using Eq.({\rm A}.7) and making some algebraic transformations with the
Dirac matrices we get
\begin{eqnarray}\label{label5.6}
\hspace{-0.5in}&&\sqrt{2E_{\Lambda^+_c}V\,2E_p V}\,<p(q)
|\bar{u}(0)\,\gamma_{\mu}(1-\gamma^5)\,c(0)|\Lambda^+_c(Q)>= -\,g_{\rm \pi
NN}\,\Bigg[\frac{4}{5}\,\frac{g_{\rm C}}{ g_{\rm
B}}\,\frac{F_{\pi}\Lambda_{\chi}}{m^3}\Bigg]\nonumber\\
\hspace{-0.5in}&&\times \,\bar{u}_p(q,\sigma^{\prime}\,)
(2\hat{v}\gamma_{\mu}\hat{q} + \hat{v}\hat{q}\gamma_{\mu})(1 -
\gamma^5)\,u_{\Lambda^+_c}(Q,\sigma).
\end{eqnarray}
The amplitude of the $\Lambda^+_c \to p + \bar{K}^0$ decay is defined
\begin{eqnarray}\label{label5.7}
&&{\cal M}(\Lambda^+_c(Q) \to p(q) K^0(q_0)) = i\,G_F\,V^*_{c s}\,V_{u
d}\,\bar{C}_2(\Lambda_{\chi})\,\frac{F_K}{m}\,\Bigg[g_{\rm \pi
NN}\,\frac{4}{5}\,\frac{g_{\rm C}}{ g_{\rm
B}}\,\frac{F_{\pi}\Lambda_{\chi}}{m^2}\Bigg]\,M^2_{\Lambda^+_c}\nonumber\\
&&\times\,\bar{u}_p(q,\sigma^{\prime}\,)\,(A +
B\,\gamma^5)\,u_{\Lambda^+_c}(Q,\sigma),
\end{eqnarray}
where the constants $A$ and $B$ are given by
\begin{eqnarray}\label{label5.8}
A &=& 1 + \frac{M_p}{M_{\Lambda^+_c}} - 2\,\frac{M^2_p}{M^2_{\Lambda^+_c}}
- \frac{M^2_{\bar{K}^0}}{M^2_{\Lambda^+_c}} = 1.03,\nonumber\\
B &=& 1 - \frac{M_p}{M_{\Lambda^+_c}} - 2\,\frac{M^2_p}{M^2_{\Lambda^+_c}}
- \frac{M^2_{\bar{K}^0}}{M^2_{\Lambda^+_c}} = 0.21.
\end{eqnarray}
The numerical values are obtained for $M_{\Lambda^+_c} =2285\,{\rm MeV}$,
$M_p = 938\,{\rm MeV}$ and $M_{\bar{K}^0} = 498\,{\rm MeV}$ [2]. The
partial width of the $\Lambda^+_c \to p + \bar{K}^0$ decay reads
\begin{eqnarray}\label{label5.9}
&&\Gamma(\Lambda^+_c \to p\bar{K}^0) = |G_F\,V^*_{c s}\,V_{u
d}\,\bar{C}_2(\Lambda_{\chi})|^2\Bigg[g_{\rm \pi
NN}\,\frac{4}{5}\,\frac{g_{\rm C}}{ g_{\rm
B}}\,\frac{F_{\pi}\Lambda_{\chi}}{m^2}\Bigg]^2\,\frac{F^2_K}{m^2}\,
\frac{M^3_{\Lambda^+_c}}{32\pi}\nonumber\\
&&\times\,{\rm tr}\{(\hat{Q} + M_{\Lambda^+_c})(A -  B\,\gamma^5)(\hat{q} +
M_p)(A +  B\,\gamma^5)\}\nonumber\\
&&\times\,\sqrt{\displaystyle \Bigg[1 - \Bigg(\frac{M_p -
M_{\bar{K}^0}}{M_{\Lambda^+_c}}\Bigg)^2\Bigg]\Bigg[1 - \Bigg(\frac{M_p +
M_{\bar{K}^0}}{M_{\Lambda^+_c}}\Bigg)^2\Bigg]}.
\end{eqnarray}
When computing  the trace over Dirac matrices we obtain the partial width
in the form
\begin{eqnarray}\label{label5.10}
\hspace{-0.5in}&&\Gamma(\Lambda^+_c \to p\bar{K}^0) = |G_F\,V^*_{c s}\,V_{u
d}\,\bar{C}_2(\Lambda_{\chi})|^2\Bigg[g_{\rm \pi
NN}\,\frac{4}{5}\,\frac{g_{\rm C}}{ g_{\rm
B}}\,\frac{F_{\pi}\Lambda_{\chi}}{m^2}\Bigg]^2 \Bigg[\frac{5
M^5_{\Lambda^+_c}}{512\pi^3}\Bigg]\nonumber\\
\hspace{-0.5in}&&\times\,\frac{32\pi^2}{5}\,\frac{F^2_K}{m^2}\,
\Bigg\{A^2\,\Bigg[\Bigg(1 + \frac{M_p}{M_{\Lambda^+_c}}\Bigg)^2 -
\frac{M^2_{\bar{K}^0}}{M^2_{\Lambda^+_c}}\Bigg] + B^2\,\Bigg[\Bigg(1 -
\frac{M_p}{M_{\Lambda^+_c}}\Bigg)^2 -
\frac{M^2_{\bar{K}^0}}{M^2_{\Lambda^+_c}}\Bigg]\Bigg\}\nonumber\\
\hspace{-0.5in}&&\times\,\sqrt{\displaystyle \Bigg[1 - \Bigg(\frac{M_p -
M_{\bar{K}^0}}{M_{\Lambda^+_c}}\Bigg)^2\Bigg]\Bigg[1 - \Bigg(\frac{M_p +
M_{\bar{K}^0}}{M_{\Lambda^+_c}}\Bigg)^2\Bigg]}=\nonumber\\
\hspace{-0.5in}&&= 11.72\,\times\,|G_F\,V^*_{c s}\,V_{u
d}\,\bar{C}_2(\Lambda_{\chi})|^2\Bigg[g_{\rm \pi
NN}\,\frac{4}{5}\,\frac{g_{\rm C}}{ g_{\rm
B}}\,\frac{F_{\pi}\Lambda_{\chi}}{m^2}\Bigg]^2 \Bigg[\frac{5
M^5_{\Lambda^+_c}}{512\pi^3}\Bigg].
\end{eqnarray}
Now we can define the partial width of the $\Lambda^+_c \to p + \bar{K}^0$
decay with respect to the partial width of the $\Lambda^+_c \to p + K^- +
\pi^+$ decay. By using Eqs.(\ref{label4.9})  and (\ref{label4.10}) we get
\begin{eqnarray}\label{label5.11}
\hspace{-0.5in}R_{\rm th} = \frac{\displaystyle \Gamma(\Lambda^+_c \to
p\bar{K}^0)}{\displaystyle \Gamma(\Lambda^+_c \to pK^-\pi^+)} =
180.31\,\times\,\frac{\bar{C}^2_2(\Lambda_{\chi})}{\bar{C}^2_1(\Lambda_{\chi})}.
\end{eqnarray}
The numerical factor in front of the ratio of the squared Wilson
coefficients is completely due the low---energy dynamics of our model
induced by a linearly rising confinement potential. In order to verify the
consistency of this dynamics with a short--distance QCD we should
substitute in Eq.(\ref{label5.11}) the numerical values of the Wilson
coefficients.

Following Buras et al. [15] we obtain $C_1(\Lambda_{\chi}) = 1.24$  and
$C_2(\Lambda_{\chi}) = - 0.47$. These numerical values agree well with the
numerical values of the Wilson coefficients calculated at the normalization
scale $\mu \simeq 1.5\,{\rm GeV}$ [15]: $C_1(\mu) \simeq 1.21$  and
$C_2(\mu) \simeq - 0.42$.

The ratio $R_{\rm th}$ calculated at $C_1(\Lambda_{\chi}) = 1.24$  and
$C_2(\Lambda_{\chi}) = - 0.47$ amounts to
\begin{eqnarray}\label{label5.12}
R_{\rm th} = \frac{\displaystyle \Gamma(\Lambda^+_c \to
p\bar{K}^0)}{\displaystyle \Gamma(\Lambda^+_c \to pK^-\pi^+)} = 0.50.
\end{eqnarray}
The theoretical result agrees  well with the experimental value averaged
over all experimental data [2]: $R_{\exp}=(0.49\pm 0.07)$. This agreement
is non--trivial and testifies not only the consistency of the  model with
short--distance QCD but the self--consistency of the effective quark model
with chiral $U(3)\times U(3)$ symmetry incorporating HQET and (CHPT)$_q$.
The former is due to the distinction between the computations of the matrix
elements of the modes  $\Lambda^+_c \to p + \bar{K}^0$ and $\Lambda^+_c \to
p + K^- + \pi^+$. Indeed, the computation of the matrix element of the mode
$\Lambda^+_c \to p + \bar{K}^0$  admits the factorization of the baryonic
and mesonic degrees of freedom, whereas for the computation of the matrix
element of the mode $\Lambda^+_c \to p + K^- + \pi^+$ such a factorization
is not feasible.

Our theoretical result for ratio Eq.(\ref{label5.12}) confirms too the
validity of our prediction for the proton energy spectrum and the angular
distribution of the probability of the  $\Lambda^+_c \to p + K^- + \pi^+$
decay given by Eq.(\ref{label4.12}) and Eq.(\ref{label4.13}), respectively.

\section{Conclusion}

The main result of the paper is in the prediction of the proton energy
spectrum and the angular distribution of the probability of the mode
$\Lambda^+_c \to p + K^- + \pi^+$ of the $\Lambda^+_c$ decays. This mode is
the most favourable for the measurement as it contains the proton and the
charged mesons. However, from the theoretical point of view this mode is
the most difficult for the computation  due to the impossibility to
factorize  baryonic and mesonic degrees of freedom.

To the computation of the matrix element of the $\Lambda^+_c \to p + K^- +
\pi^+$ decay we have applied the effective quark model with chiral
$U(3)\times U(3)$ symmetry incorporating HQET and (CHPT)$_q$. This model is
motivated by the effective low--energy QCD with a linearly rising
confinement potential. Due to the dynamics of strong low--energy
interactions caused by a linearly rising confinement potential the
spinorial structure of the three--quark currents coupled to the baryons is
fixed unambiguously. The effective low--energy interactions of the
low--lying baryon octet and charmed baryons coupled to the three--quark
currents can be described in terms of two phenomenological coupling
constants $g_{\rm B}$ and $g_{\rm C}$, respectively. These constants enter
multiplicatively to the matrix elements of the strong low--energy
transitions of baryons.

In the case of the non--leptonic decays of the $\Lambda^+_c$ the
multiplicative character of the constants $g_{\rm B}$ and $g_{\rm C}$
allows to replace the product of these constants by the experimental value
of the probability of the mode $\Lambda^+_c \to p + K^- + \pi^+$. This
defines any mode of the $\Lambda^+_c$ decays relative to the mode
$\Lambda^+_c \to p + K^- + \pi^+$. As regards the proton energy spectrum
and the angular distribution, the resultant expressions do not contain any
arbitrary parameters and can be applied to the analysis of the polarization
of the $\Lambda^+_c$ in the processes of photo and hadroproduction. We have
considered the simplest case, maybe most favourable from the experimental
point of view, when the $\Lambda^+_c$ is polarized while the proton of the
decay is unpolarized. This means that for the restoration of the
polarization the $\Lambda^+_c$ one should follow only the geometry of the
momenta of the protons of the decay but not their polarizations. In this
case there is an obvious similarity between measurements of the
polarization of the $\Lambda^+_c$ in the $\Lambda^+_c \to p + K^- + \pi^+$
decay and the $\mu^-$--meson in the $\beta$--decay $\mu^- \to e^- +
\bar{\nu}_{\rm e} + \nu_{\mu}$.

For the confirmation of the validity of our prediction for the proton
energy spectrum and the angular distribution of the probability of the mode
$\Lambda^+_c \to p + K^- + \pi^+$, we have computed the probability of the
mode $\Lambda^+_c \to p + \bar{K}^0$ relative to the probability of the
mode $\Lambda^+_c \to p + K^- + \pi^+$. Our prediction for the ratio of the
probabilities $R_{\rm th} = B(\Lambda^+_C \to p\bar{K}^0)/B(\Lambda^+_c \to
p K^- \pi^+) = 0.50$ agrees well with the experimental value averaged over
all experimental data $R_{\exp}=(0.49\pm 0.07)$.  This agreement is not
trivial and confirms not only the self--consistency of our approach but the
consistency of it with a short--distance QCD, since the computation of the
matrix element of the mode $\Lambda^+_c \to p + \bar{K}^0$ differs from the
computation of the matrix element of the mode $\Lambda^+_c \to p + K^- +
\pi^+$. Indeed, if for the computation of the matrix element of the mode
$\Lambda^+_c \to p + \bar{K}^0$ one can factorize the baryonic and mesonic
degrees of freedom, whereas in the case of the computation of the matrix
element of the mode $\Lambda^+_c \to p + K^- + \pi^+$ such a factorization
is not feasible.

\newpage

\section*{Appendix. Computation of the momentum integrals}

We perform the integration over $k_1$ and $k_2$ of the momentum integral of
Eq.(\ref{label3.6}). For this aim we consider the integral
$$
{\cal J}(v) = \int\frac{d^4k_1}{\pi^2i}\int
\frac{d^4k_2}{\pi^2i}\,\frac{1}{m - \hat{k}_1}\Gamma\frac{1}{m + \hat{k}_2}
\frac{1}{[(k_1 + k_2)\cdot v + i0]},\eqno({\rm A}.1)
$$
where $\Gamma$ is a Dirac matrix. The nontrivial contribution comes from
the components of the $k^{\mu}_1$ and $k^{\mu}_2$ 4--vectors parallel to
4--vector $v^{\mu}$. This gives
$$
{\cal J}(v) = \int\frac{d^4k_1}{\pi^2i}\int \frac{d^4k_2}{\pi^2i}\,\frac{m
+ \hat{v} k_1\cdot v}{[m^2 - k^2_1 - i0]}\Gamma\frac{m - \hat{v} k_2\cdot
v}{[m^2 - k^2_2 - i0]}\frac{1}{[(k_1 + k_2)\cdot v + i0]}.\eqno({\rm A}.2)
$$
Now it is convenient to make the Wick rotation and pass to Euclidean
momentum space [9]:
$$
{\cal J}(v) = 4\,i\int\limits^{\infty}_0\frac{dk_{E1} k_{E1}^3}{m^2 +
k^2_{E1}}\int \frac{d\Omega_1}{2\pi^2}(m +i\hat{v}\,k_{E1}\cos\chi_1)
\Gamma \int\limits^{\infty}_0\frac{dk_{E2} k_{E2}^3}{m^2 + k^2_{E2}}
$$
$$
\int \frac{d\Omega_2}{2\pi^2}\,\frac{m -i\hat{v}\,
k_{E2}\cos\chi_2}{k_{E1}\cos\chi_1 + k_{E2}\cos\chi_2},\eqno({\rm A}.3)
$$
where $d\Omega_i=4\pi\sin^2\chi_i d\chi_i\,(i=1,2)$ are the solid angles in
Euclidean spaces of the momenta $k^{\mu}_{E1}$ and $k^{\mu}_{E2}$,
respectively, $k_{Ei} =\sqrt{k^2_{4i} + \vec{k}^{\,2}_i}\,(i=1,2)$. Then we
have used the relation $\sqrt{v^2_E} = - i$ [8].

In order to disconnect integrations over $k_{E1}$ and $k_{E2}$ we suggest
to use the following integral representation
$$
{\cal J}(v) = 4 \int\limits^{\infty}_0 dt
\int\limits^{\infty}_0\frac{dk_{E1} k_{E1}^3}{m^2 + k^2_{E1}}\int
\frac{d\Omega_1}{2\pi^2}(m +i\hat{v} \,k_{E1}\cos\chi_1) \,e^{\textstyle
i\,t\,k_{E1}\cos\chi_1}\,\Gamma
$$
$$
\int\limits^{\infty}_0\frac{dk_{E2}k_{E2}^3}{m^2 + k^2_{E2}}
\int \frac{d\Omega_2}{2\pi^2}(m -i\hat{v}\,
k_{E2}\cos\chi_2)\,\,e^{\textstyle i\,t\, k_{E2} \cos\chi_2}.\eqno({\rm
A}.4)
$$
Integrating out $\chi_1$ and $\chi_2$ we get
$$
{\cal J}(v) = 16 \int\limits^{\infty}_0 dt
\int\limits^{\infty}_0\frac{dk_{E1} k_{E1}^3}{m^2 +
k^2_{E1}}\Bigg\{m\,\Bigg[\frac{J_1(k_{E1}t)}{k_{E1}t}\Bigg] -
i\,\hat{v}\,\frac{\partial}{\partial
t}\Bigg[\frac{J_1(k_{E1}t)}{k_{E1}t}\Bigg]\Bigg\}\,\Gamma
$$
$$
\int\limits^{\infty}_0\frac{dk_{E2}k_{E2}^3}{m^2 +
k^2_{E2}}\Bigg\{m\,\Bigg[\frac{J_1(k_{E2}t)}{k_{E2}t}\Bigg] +
i\hat{v}\,\frac{\partial}{\partial
t}\Bigg[\frac{J_1(k_{E2}t)}{k_{E2}t}\Bigg]\Bigg\},\eqno({\rm A}.5)
$$
where $J_1(k_{Ei}t)\,(i=1,2)$ is the Bessel function. Now we can perform
the integration over $k_{Ei}\,(i=1,2)$:
$$
{\cal J}(v) = 16 m^2 \int\limits^{\infty}_0
dt\Bigg\{m\,\Bigg[\frac{K_1(mt)}{t}\Bigg] -
i\hat{v}\,\frac{\partial}{\partial
t}\Bigg[\frac{K_1(mt)}{t}\Bigg]\Bigg\}\Gamma
\Bigg\{m\,\Bigg[\frac{K_1(mt)}{t}\Bigg] +
i\hat{v}\,\frac{\partial}{\partial
t}\Bigg[\frac{K_1(mt)}{t}\Bigg]\Bigg\},\eqno({\rm A}.6)
$$
where $K_1(mt)$ is the McDonald function. The integrals over $t$ are
divergent and should be regularized. We suggest to use the cut--off
regularization restricting the region of the integration over $t$ from
below as $t\ge 1/\Lambda_{\chi}$. Keeping leading divergent contributions
[5--12] it is convenient to represent the r.h.s. of Eq.({\rm A}.6) in the
following form:
$$
{\cal J}(v) =
\frac{4}{5}\,\Bigg[\frac{16\pi^2}{3}\Bigg]^2\,\frac{\Lambda_{\chi}}{m^2}\,<\bar{
q}q>^2\,\hat{v}\,\Gamma\,\hat{v},\eqno({\rm A}.7)
$$
where $<\bar{q}q>$ is the quark condensate  defined in terms of the
SB$\chi$S scale and the constituent quark mass as follows [5--12]
$$
<\bar{q}q> = - \frac{N}{16\pi^2}\int\frac{d^4k}{\pi^2i}{\rm
tr}\Bigg\{\frac{1}{m - \hat{k}}\Bigg\}=- \frac{N
m}{4\pi^2}\Bigg[\Lambda^2_{\chi} - m^2{\ell n}\Bigg(1 +
\frac{\Lambda^2_{\chi}}{m^2}\Bigg)\Bigg] = - (253\,{\rm MeV})^3. \eqno({\rm
A}.8)
$$
The numerical value is calculated at $N = 3$, $m = 330\,{\rm MeV}$ and
$\Lambda_{\chi} = 940\,{\rm MeV}$. As has been shown in Ref.[5] the quark
condensate value $<\bar{q}q> = - (253\,{\rm MeV})^3$ describes with an
accuracy better than 5$\%$ the mass spectrum of low--lying pseudoscalar
mesons for the current quark masses $m_{0 u} = 4\,{\rm MeV}$, $m_{0 d} =
7\,{\rm MeV}$ and $m_{0 s} = 135\,{\rm MeV}$ quoted by QCD [18].

\end{document}